\documentclass[pre,aps,twocolumn]{revtex4}

\usepackage{times}
\usepackage{amsmath}
\usepackage{graphicx}
\usepackage{amssymb}
\usepackage{color}
\usepackage{epstopdf}

\newcommand{\e}{\mathrm{e}}

\newcommand{\beq}{\begin{equation}}
\newcommand{\eeq}{\end{equation}}
\newtheorem{remark}{Remark}

\begin{document}

 \title{A Unifying Perspective: Solitary Traveling Waves As Discrete Breathers \\
  in Hamiltonian Lattices and Energy Criteria for Their Stability}

\author{Jes\'us Cuevas--Maraver}
\affiliation{Grupo de F\'{\i}sica No Lineal, Departamento de F\'{\i}sica Aplicada I, Universidad de Sevilla. Escuela Polit{\'e}cnica Superior, C/ Virgen de \'Africa, 7, 41011-Sevilla, Spain \\ Instituto de Matem\'aticas de la Universidad de Sevilla (IMUS). Edificio Celestino Mutis. Avda. Reina Mercedes s/n, 41012-Sevilla, Spain}

\author{Panayotis\ G.\ Kevrekidis}
\affiliation{Department of Mathematics and Statistics, University of
Massachusetts, Amherst, MA 01003-9305, USA}

\author{Anna Vainchtein}
\affiliation{Department of Mathematics, University of Pittsburgh, Pittsburgh, PA 15260, USA}

\author{Haitao Xu}
\affiliation{Institute for Mathematics and its Applications, University of
  Minnesota,
Minneapolis, MN 55455 USA}

\begin{abstract}
  In this work, we provide two complementary perspectives for the
  (spectral) stability of solitary traveling waves in Hamiltonian nonlinear
  dynamical lattices, of which the Fermi-Pasta-Ulam and the Toda
  lattice are prototypical examples. One is as an eigenvalue problem
  for a stationary solution in a co-traveling frame, while the other
  is as a periodic orbit modulo shifts. We connect the eigenvalues of
  the former with the Floquet multipliers of the latter and based on
  this formulation  derive an energy-based spectral stability
  criterion. It states
  that a sufficient (but not necessary) condition for a
  change in the wave stability occurs when the functional dependence of the energy (Hamiltonian) $H$ of
  the model on the wave velocity $c$ changes its monotonicity.
  Moreover, near the critical velocity where the change of stability occurs,
      we provide an explicit leading-order computation of the unstable eigenvalues, based on the second derivative of the Hamiltonian $H''(c_0)$
      evaluated at the critical velocity $c_0$.
  We corroborate this conclusion
  with a series of analytically and numerically tractable examples
  and discuss its parallels with a recent energy-based criterion for
  the stability of discrete breathers.
\end{abstract}

\maketitle

\section{Introduction} 
Solitary traveling waves (STWs) are ubiquitous in Hamiltonian lattice dynamical
systems with intersite interactions. They arise
in the model at the very foundation of nonlinear science, namely
the Fermi-Pasta-Ulam (FPU) lattice~\cite{FPU}, as well as in the Toda lattice~\cite{toda},
one of the key systems of interacting particles, and, arguably,
the most significant integrable one.
In addition to their theoretical relevance in the above models,
they constitute the most generic, robust and often experimentally
tractable excitation in nonlinear systems, in particular, in
granular crystals~\cite{nesterenko,sen,review} and other materials.

Given the relevance of STWs in
theoretical, numerical \cite{mertens,eilbeck,english} and experimental \cite{nesterenko,sen} studies,
it is natural to be concerned about their
stability. This may be accessible in some special cases, such as
the Toda lattice~\cite{wayne}, or the FPU problem in the low-energy (near-sonic) regime~\cite{pegof3,pegof4}, where
specialized techniques become available due to the system's  integrability (or proximity to it). Nevertheless, from a physical perspective, it would be
desirable to have a more general criterion that would be intuitive
as well as straightforward to test. This is especially important given
that in a number of studies~\cite{mertens1,mertens2,at}, the possibility
of unstable STWs has been demonstrated.

In the present work, we offer such a criterion (a sufficient
yet not necessary condition) by establishing
that a change in the monotonicity of the STW's energy (Hamiltonian $H$)
dependence on the velocity $c$ will result in a change in its (spectral)
stability. In other words, we establish that when,
for a critical velocity $c_0$, it happens that $H'(c_0)=0$, a pair of eigenvalues
associated with the traveling wave vanish, entailing the potential
for instability. While this criterion first appeared in~\cite{pegof3}, where it was
motivated by the study of the FPU problem in the near-sonic limit,
here we provide both a concise proof, and also a definitive
leading-order calculation for these two near-zero eigenvalues to explicitly show why (and when) instability appears.
We also
systematically test
the criterion numerically in a broad array of physically relevant cases.

Equally important in our approach is the fact that we provide a
generalized perspective of the problem of the stability of STWs in a
Hamiltonian lattice. In the frame traveling with the solution, the stability leads to a standard
eigenvalue problem. Yet, here, motivated by earlier works such
as~\cite{floria}, we also propose a complementary approach, where the solution is viewed
as a {\it periodic
  orbit} of the map involving (a) running the solution for a period
of $h/c$, where $h$ is the lattice spacing, rescaled to unity below
and (b) shifting back by one lattice site. In light of
this periodicity, Floquet analysis can be brought to bear
and will turn out to yield {\it coincident} stability conclusions
about instabilities produced by the criterion put forth. Furthermore,
this perspective enables a unification of the lattice STWs in such Hamiltonian systems through their consideration
as discrete breathers. Here the effective frequency $\omega$ is
proportional to their velocity $c$ according to $\omega=2 \pi c/h$. This,
in turn, directly connects the criterion we analyze with a recently
established criterion for the spectral stability of discrete
breathers~\cite{dmp}. We emphasize here that
  the unifying connection of STWs with breathers does not impose any a-priori restrictions
  on the nature of their decay of at infinity.
  
The paper is organized as follows. In Sec.~\ref{sec:criterion} we formulate the problem, analyze the properties of the linear operator associated with a STW and prove the energy-based stability criterion. We also describe the behavior of the relevant eigenvalues near the critical velocity, based on the derivation presented in Appendix~\ref{sec:details}. In Sec.~\ref{sec:floquet} we discuss an alternative perspective for the spectral stability, which is associated with the Floquet analysis. Our results are corroborated by numerical examples in Sec.~\ref{sec:examples}, with further details provided in Appendix~\ref{sec:numer}. We summarize our findings and discuss some open questions in Sec.~\ref{sec:conclusions}.

\section{Stability Analysis in the Co-traveling Frame and the
  Energy Criterion}
\label{sec:criterion} 
We consider a rescaled Hamiltonian system of
the form
\begin{equation}
\dfrac{du}{dt} = p, \quad \dfrac{dp}{dt} = F(u)=-\frac{\partial {\cal H}}{\partial u},
  \label{eqn_main}
\end{equation}
where ${\cal H}$ denotes the Hamiltonian energy
density of the system, $u(t)$ and $p(t)$ are infinite-dimensional vectors denoting the displacement and particle velocity values on the lattice, with components $u_n$ and $p_n$, respectively. In a more compact notation, Eq.~\eqref{eqn_main} can be written as
\beq
\dfrac{dU}{dt}= J \nabla {\cal H}(U),
\label{eq0}
\eeq
where
\[
U=\left(\begin{array}{c}
  {u} \\
  {p}
\end{array} \right), \quad J=\left(\begin{array}{cc}
0 & I \\
-I & 0
\end{array} \right).
\]
We assume the existence of STWs for a continuous interval of velocities. These are localized solutions of the
form 
\[
u_n(t)=\hat{u}(\xi), \quad p_n(t)=\hat{p}(\xi), \quad \xi=n-ct, 
\]
where $c$ denotes the velocity of the wave, and $\xi$ is
the co-traveling frame variable (note that $\hat{p}(\xi)=-c \hat{u}'(\xi)$), with finite energy (see Appendix~\ref{sec:details} for more details).
Linearization about the STW in the co-traveling frame, with $u(\xi,t)=\hat{u}(\xi)+ \epsilon e^{\lambda t}
W(\xi)$ and $p(\xi,t)=\hat{p}(\xi) + \epsilon e^{\lambda t} P(\xi)$ for small $\epsilon$, then yields
the eigenvalue problem
\beq
\label{eq_ev}
\lambda Z=\mathcal{L}Z
\eeq
for the linear operator
\beq
\mathcal{L}= c \partial_{\xi}+J \nabla^2 {\cal H}(\hat{U}),
\label{eq:L}
\eeq
where
\[
Z=\left(\begin{array}{c}
  {W} \\ {P}
  \end{array} \right), \quad
\hat{U}=\left(\begin{array}{c}
  {\hat{u}} \\ {\hat{p}}
  \end{array} \right), \quad
J \nabla^2 \mathcal{H}(\hat{U})=\left(\begin{array}{cc}
0 & I \\
F'(\hat{u}) & 0
\end{array} \right).
\]
Solving the problem in Eq.~\eqref{eq_ev} provides information about the stability of the STW, through the
spectrum of the linearization operator $\mathcal{L}$, with adjoint
\beq
\mathcal{L}^*=(- \nabla^2 {\cal H}(\hat{U}) J- c \partial_{\xi})=-J^{-1}\mathcal{L}J
\label{eq:Lstar}
\eeq
(note that $J\mathcal{L}$ is self-adjoint).
Given the time translation symmetry, an important feature of
$\mathcal{L}$ is the existence of an eigenvector $e_0=-\partial_{\xi}\hat{U}$ associated with eigenvalue $\lambda=0$. The corresponding
generalized eigenfunction is $e_1=\partial_{c} \hat{U}$, i.e., $\mathcal{L}e_1=e_0$.
In other words, the spectrum of $\mathcal{L}$ always contains a double eigenvalue at zero. Moreover, by symmetry,
  the algebraic multiplicity of the zero eigenvalue can only be even.

The presence of an additional instability presupposes the {\it increase}
of the algebraic multiplicity of the $0$ eigenvalue. Since the kernel of $\mathcal{L}$ is one-dimensional, an algebraic multiplicity higher than two (i.e., at least four) implies that there exists $e_2$ such that
$\mathcal{L}e_2 = e_1=\partial_{c} \hat{U}$. Since $J^{-1}e_0=-J^{-1}\partial_{\xi}\hat{U}$ is in the kernel of $\mathcal{L}^*$, this
yields the solvability condition
\[
\begin{split}
0 &= \langle  J^{-1}e_0,e_1 \rangle =\int (-J^{-1}\partial_{\xi} \hat{U}) \cdot (\partial_{c}\hat{U}) d \xi \\
&= \int \frac{1}{c} \nabla {\cal H} (\hat{U})  \cdot \frac{\partial \hat{U}}{\partial c} d\xi
= \frac{1}{c}\int \frac{\partial {\cal H}(\hat{U})}{\partial c}  d \xi = \frac{1}{c} H'(c),
\end{split}
\]
where $H=\int  {\cal H} d \xi$ is the conserved Hamiltonian of the system,
and $\langle \cdot , \cdot \rangle$ denotes the relevant inner product.

As soon as $c$ deviates from the critical velocity $c_0$
satisfying $H'(c_0)=0$, the above solvability condition fails (e.g. assuming $H''(c_0)\neq 0$), and hence two eigenvalues start to move away from zero and can possibly
emerge on the real axis. Thus the condition  $H'(c_0)=0$ constitutes a
threshold for instability of STWs, as per the concise
proof above and detailed numerical considerations below extending
the formulation of~\cite{pegof3}. In fact, by computing the leading-order approximation of these two near-zero eigenvalues near $c_0$ one can reveal the trend of their motion.  Suppose, as will be typically the case when the stability changes
  (including examples in Sec.~\ref{sec:examples}
  below), that the generalized kernel of $\mathcal{L}$ is exactly four-dimensional at $c_0$, with
$\mathcal{L}^3e_3=\mathcal{L}^2e_2=\mathcal{L}e_1=e_0=-\partial_{\xi}\hat{U}$.
  Then, as shown in Appendix~\ref{sec:details}, the pair of eigenvalues of $\mathcal{L}$ responsible for
  the change of stability will be given by
\begin{equation}
\label{eq_ev_pred}
 \lambda=\pm \sqrt{\frac{H''(c_0)}{\alpha_1 c_0}(c-c_0)}+O(|c-c_0|) 
\end{equation}
for $c$ is near $c_0$, where nonzero $\alpha_1$ is defined in \eqref{eq:alpha1} in terms of generalized eigenvectors.

In Sec.~\ref{sec:examples} we numerically verify the theoretical predictions (and test the validity of Eq.~\eqref{eq_ev_pred}),
showing that a change of the monotonicity of $H(c)$ will constitute a sufficient (but not
necessary) condition for the transition from stability to instability,
or vice versa, depending on the sign of $H''(c_0)\alpha_1c_0$.

\section{A Complementary Perspective: Floquet Analysis of the
  time $T=h/c$ map}
\label{sec:floquet}    
Let us now envision anew the case of
a STW on a lattice. Over the period $T=h/c$ (below we again
set $h=1$), the STW $\hat{U}$ moves over
by one lattice site. However, due to the integer shift
invariance of the lattice, the configuration has to be
identical to the one with which we started.
This means that upon running for a period and shifting back
using the shift operator $S$ such that $Su_n(t)=u_{n-1}(t)$,
we generate a {\it periodic orbit} on the lattice~\cite{floria}.
Thus, a fixed point of this operation consisting of
(a) run for $T=1/c$ and (b) shift, is a discrete breather (DB) i.e.,
a localized time-periodic
solution~\cite{aubry,FlachPR2008}
by construction with frequency $\omega=2 \pi c$. Yet, at the same
time the resulting profile constitutes a lattice STW.

Two important consequences of this complementary perspective are as follows.
(1) The fixed point operation discussed above has a corresponding
monodromy matrix~\cite{aubry,FlachPR2008,arnold} whose eigenvalues
are the Floquet multipliers (FMs) of the relevant periodic orbit.
These FMs determine the stability of the periodic orbit
(i.e., in this case of the STW), as do
the eigenvalues of co-traveling problem computation. Hence, one
should expect that an instability manifested through an eigenvalue
crossing zero should be accompanied by a FM $\mu$ crossing unity,
due to the well known relation $\mu=e^{\lambda T}$ between the multipliers and eigenvalues \cite{ourjesus}.
(2) Given the
intimate connection of lattice STWs and DBs, an immediate correlation emerges
between the criteria for stability change of discrete breathers,
such as $H'(\omega)=0$ that was recently established in~\cite{dmp}
and the stability of lattice STWs discussed here (and
also in~\cite{pegof3}). Observing that
for lattice STWs, $\omega=2 \pi c$, an alternative derivation of the latter
from the former is, in fact, immediate.

\section{Numerical Corroboration}
\label{sec:examples} 
We now test the above prediction in a
set of numerical examples with the generalized Hamiltonian of the form
\begin{equation}
\label{eq:Ham}
\begin{split}
    H &= \sum_{n=-\infty}^{\infty} \biggl[\frac{p_n^2}{2}+V(u_{n+1}-u_n)\\
    &+\sum_{m=-\infty}^{\infty} \frac{\Lambda(m)}{4} (u_n-u_{n+m})^2\biggr].
\end{split} 
\end{equation}
Here $V(u)$ is a generic potential governing the nonlinear interactions between nearest neighbors, and 
$\Lambda(m)$ are the coefficients of all-to-all linear long-range interactions, which decay as $|m| \rightarrow \infty$; in the absence of such interactions, $\Lambda(m)=0$. For instance, 
\beq
\Lambda(m)=\rho(\e^\gamma-1)\e^{-\gamma|m|}(1-\delta_{m,0}), 
\label{eq:KB}
\eeq
with $\rho>0$ and $\gamma>0$, corresponds to the Kac-Baker interactions, and $\Lambda(m)=\rho|m|^{-s}(1-\delta_{m,0})$ with $s=5$ ($s=3$) corresponds to the dipole-dipole (Coulomb) interactions between charged particles on a lattice.
In principle, the methodology can capture
nonlinear long-range interactions, but here we consider linear ones
for simplicity.

As our first example, we consider
the analytically tractable and well known case of the Toda lattice \cite{toda}
where $V(u)=e^{-u}+u-1$ while $\Lambda(m)=0$, which has
a one-soliton solution of the form $u_n(t)=\log\left[\cosh(\kappa(n-ct-1))\mathrm{sech}(\kappa(n-ct))\right]$,
where $\kappa$ is the unique positive solution of $c\kappa=\sinh(\kappa)$.
The resulting Hamiltonian can be computed explicitly for the
single soliton family: $H=\sinh(2\kappa)-2\kappa$, leading to $H'(c)=2(\cosh(2\kappa)-1)\partial_c\kappa>0$,
resulting in generically (spectrally)
stable solitary waves in the Toda lattice. This is also
in tune with the nonlinear stability of the solitary waves
in this case, which has been explored, e.g., in~\cite{mizu}.

\begin{figure}
\begin{center}
\begin{tabular}{cc}
\multicolumn{2}{c}{\includegraphics[width=6cm]{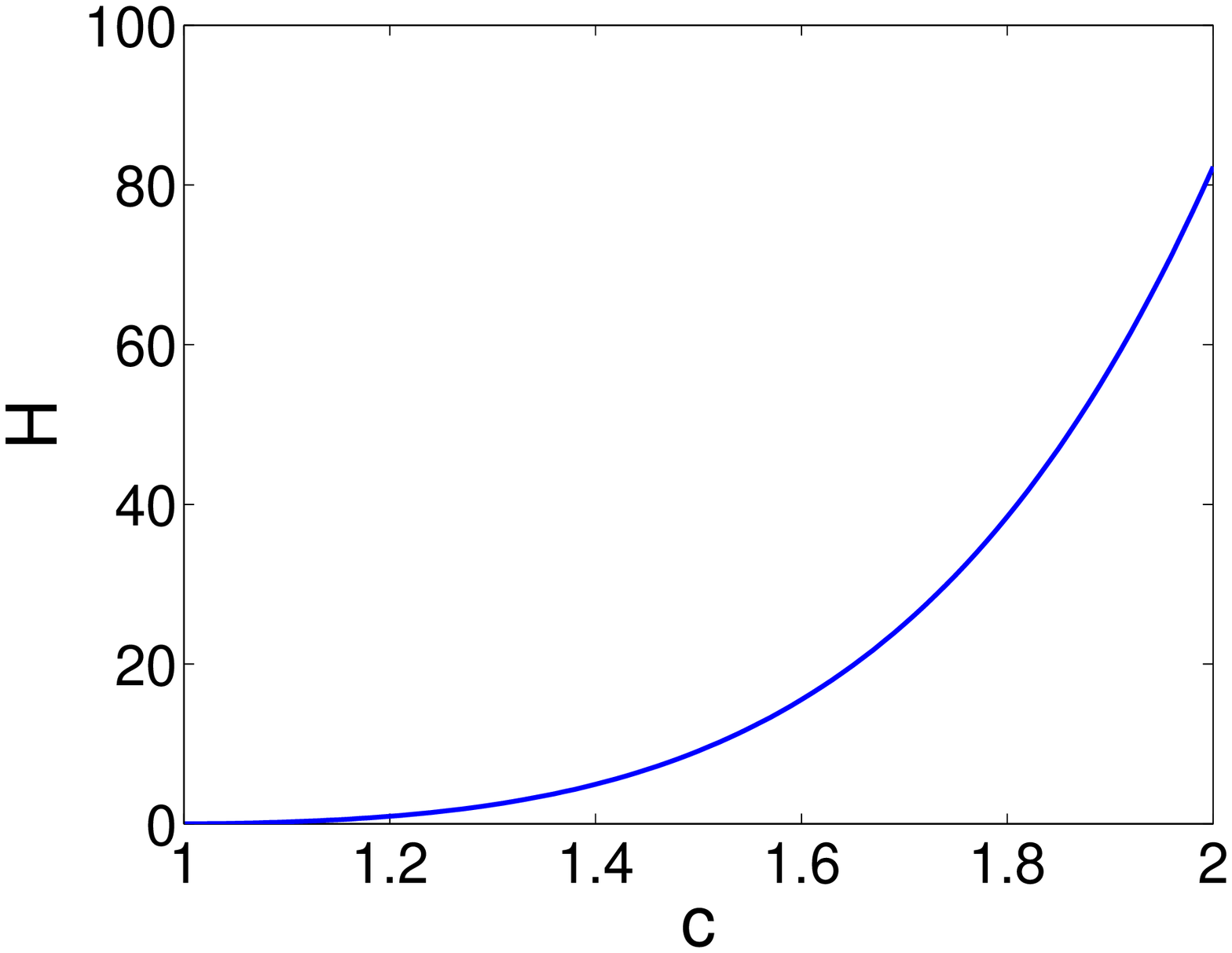}} \\
\includegraphics[width=6cm]{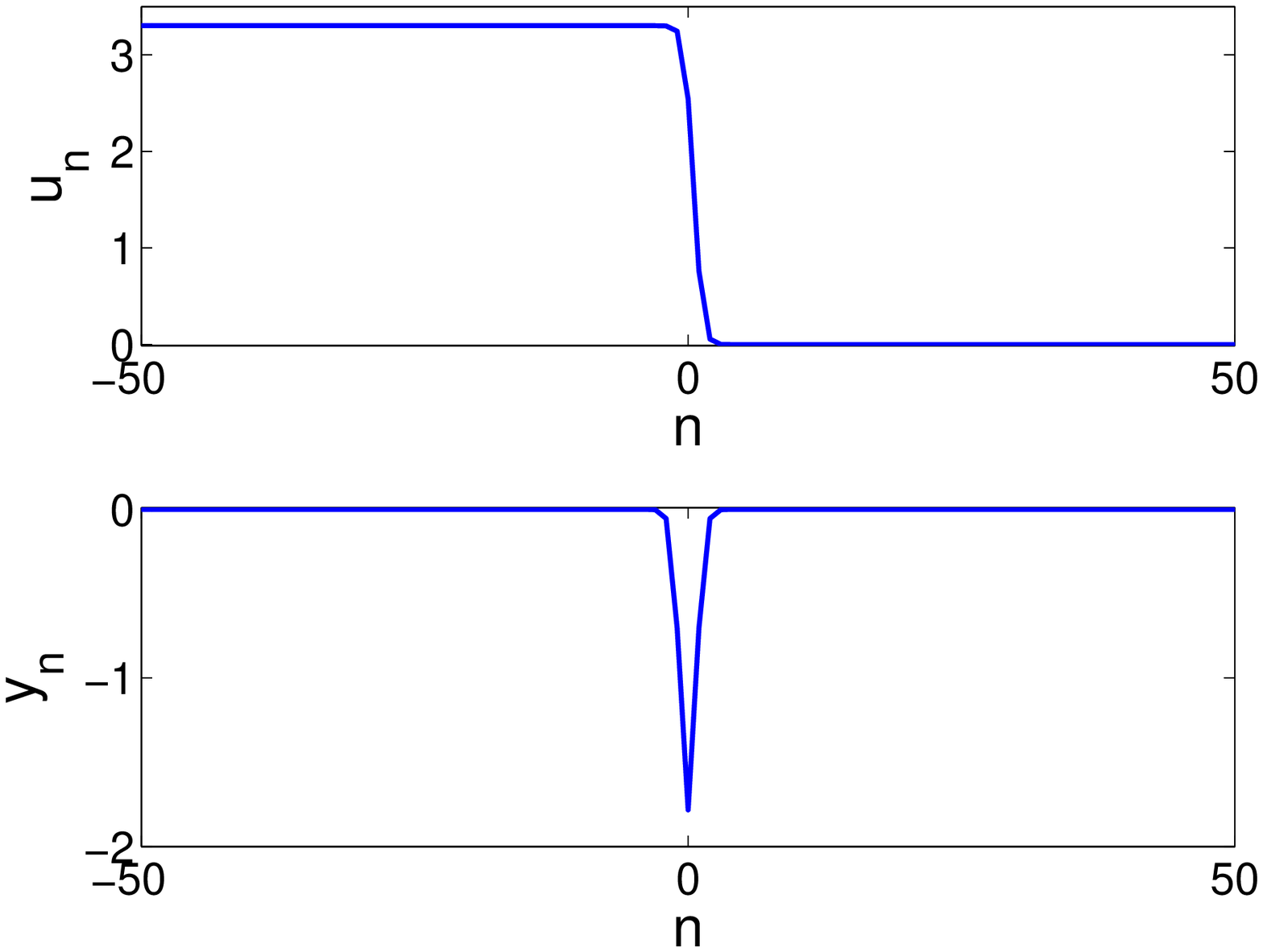} \\
\end{tabular}
\end{center}
\caption{Top panel: dependence of the energy $H$ on the wave velocity $c$ in the $\alpha$-FPU model in Eq.~\eqref{eq:FPU} with $\Lambda(m)=0$. Bottom panels: typical
  profile of the traveling wave with $c=1.5$ in the displacement ($u_n$) and strain ($y_n=u_{n+1}-u_n$) variables.}
\label{fig1}
\end{figure}

A second famous example consists of the $\alpha$-FPU case \cite{FPU}, where 
\beq
V(u)=\dfrac{u^2}{2}-\dfrac{u^3}{3},
\label{eq:FPU}
\eeq 
while $\Lambda(m)=0$.
In this case too, as identified via the methods
of~\cite{mertens,eilbeck,english,jphysa} (see Appendix~\ref{sec:numer}, for details on numerical simulations) and shown in Fig.~\ref{fig1},
the family of STWs numerically features $H'(c)>0$, in full agreement with their
identification as stable. Similar conclusions hold for the
highly experimentally relevant solitary waves of granular
crystals~\cite{nesterenko,sen,review}.
%

\begin{figure}
\begin{tabular}{c}
\includegraphics[width=8cm]{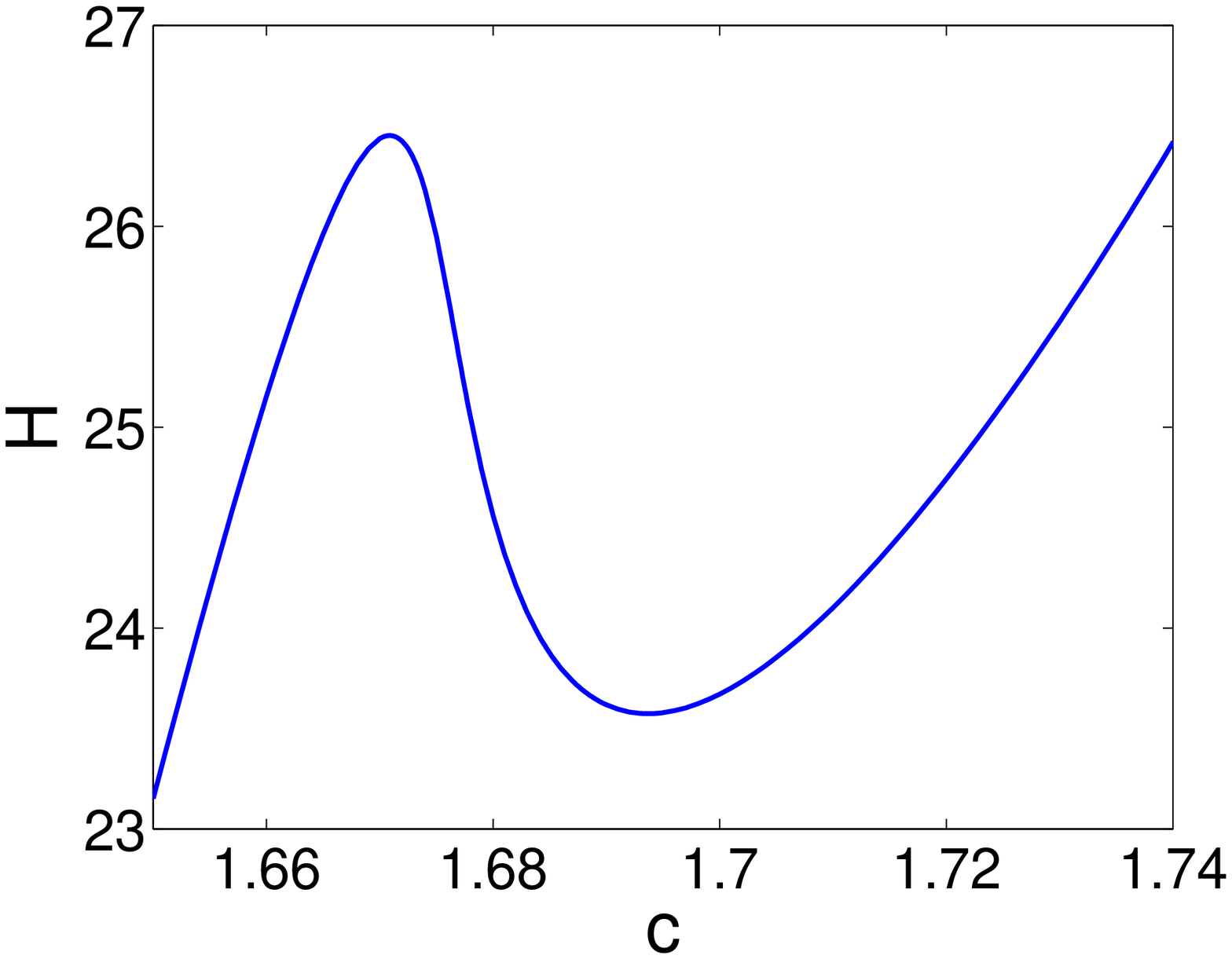} \\
\includegraphics[width=8cm]{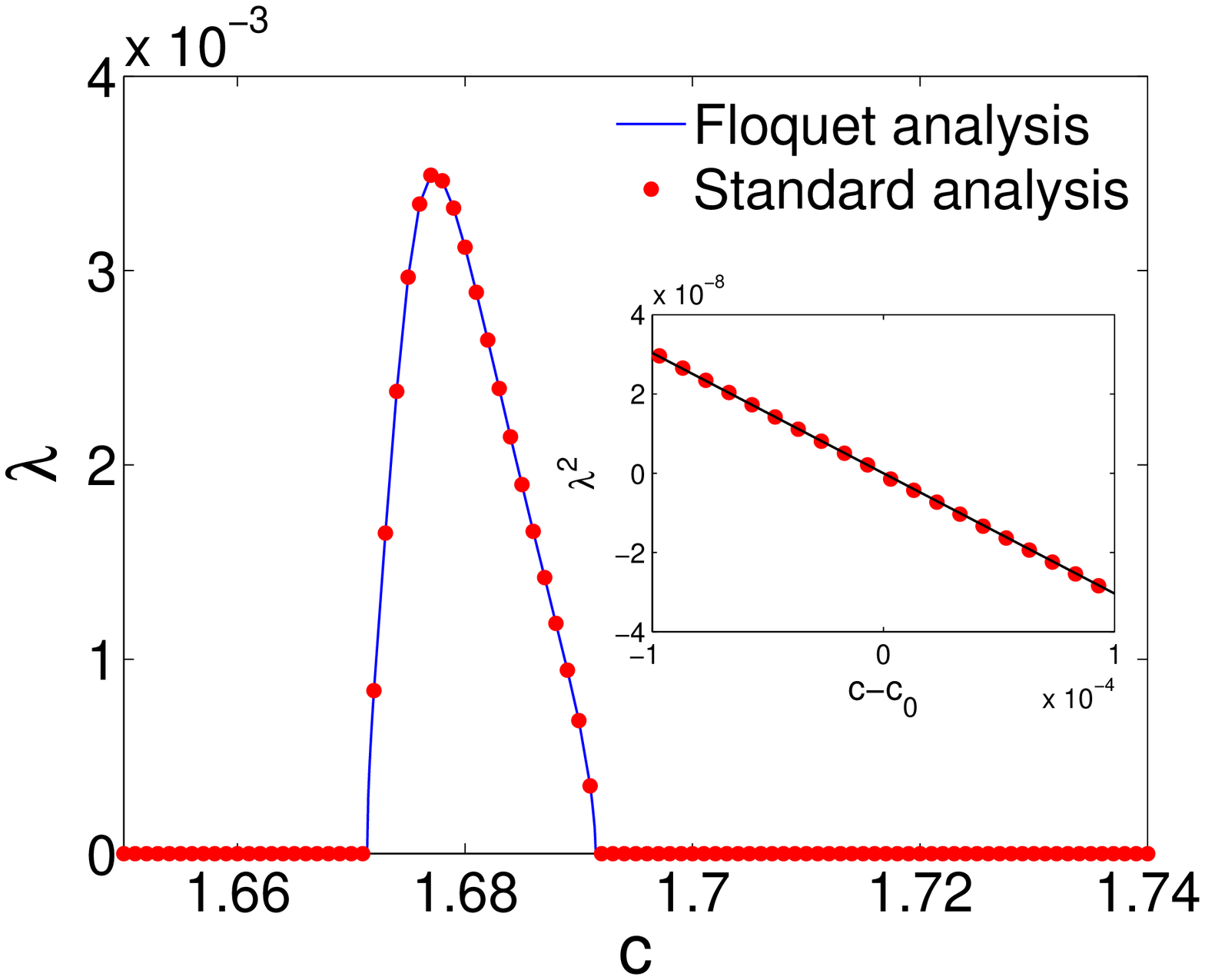} \\
\end{tabular}
\caption{Stability and instability of the lattice traveling waves in the $\alpha$-FPU lattice with nearest neighbor interactions governed by the potential in Eq.~\eqref{eq:FPU} and Kac-Baker long-range interactions with the kernel in Eq.~\eqref{eq:KB}. Here $\gamma=0.17$ and $\rho=0.0172$. The top panel shows the energy dependence on the speed, with $H'(c)>0$ implying (spectral) stability, and $H'(c)<0$ implying instability. The bottom panel confirms this by showing the relevant eigenvalue obtained by diagonalizing the linearization operator $\mathcal{L}$ (dots) and transforming the relevant Floquet multiplier $\mu$ into a corresponding eigenvalue (for comparison) via the relation $\lambda=c\log(\mu)$ (solid curve). The inset of the bottom panel shows the dependence of $\lambda^2$ on $c-c_0$ for $c$ near $c_0=1.6937$, the location of the second bifurcation; it fits a straight line $\lambda^2=\beta(c-c_0)$, with $\beta=-3.0383\times10^{-4}$.}
\label{fig2}
\end{figure}

Arguably, these cases, while interesting from the prototypically
nonlinear and experimental perspectives, are perhaps
somewhat less exciting from the point of view of our
criterion as they do not feature a stability change.
Hence, we turn to some examples which, while more
exotic from the point of view of practical applications,
have been argued to be of interest and, additionally,
feature a change of stability, which is especially relevant in the context of this work.
The first such case that we will consider concerns
the Kac-Baker interactions that have been argued
to be of relevance for modeling Coulomb interactions in DNA
molecules in~\cite{mertens2}. In this case, we maintain the potential in Eq.~\eqref{eq:FPU}
of the FPU case, but add long-range interactions with the kernel in Eq.~\eqref{eq:KB}.
Fig.~\ref{fig2} showcases the power of the stability criterion
and illustrates the complementary nature of the co-traveling
steady state and the periodic orbit FM calculation approaches.
It can be seen that $H'(c)$ becomes negative (the top panel
of Fig.~\ref{fig2}) for
$1.6709 < c < 1.6937$, for our chosen values of
$\gamma=0.17$, $\rho=0.0172$ selected in tune
with~\cite{mertens2}. For this very interval of velocities,
an eigenvalue of the operator $\mathcal{L}$ crosses
through $\lambda=0$ and acquires a positive real part (dots in the bottom panel of Fig.~\ref{fig2}).
In fact, it can be shown~\cite{pegof3} that the stability
problem in the co-traveling frame also possesses eigenvalues
$\lambda + i (2 \pi j c)$, where $j \in \mathbb{Z}$.
Finally, the solid curve in the bottom panel of the Fig.~\ref{fig2}
showcases the FM calculation associated with the
time $T=1/c$ map of the corresponding periodic orbit, transformed
(in order to compare with the steady state eigenvalue approach)
according to the relation $\lambda=\log(\mu)/T$.
Confirming the complementary picture put forth, we find
that in this case a FM pair crosses through $(1,0)$
and into the real axis for the exact same parametric
interval.

To connect with the theoretical analysis of Eq.~(\ref{eq_ev_pred}),
  the inset of Fig.~\ref{fig2}
  shows the dependence of $\lambda^2$ with respect to $c-c_0$, which, according to Eq.~(\ref{eq_ev_pred}), must be linear in the vicinity of $c_0\approx1.6937$ with the slope $\beta=H''(c_0)/(\alpha_1 c_0)=-2.9794\times10^{-4}$. Our numerical calculations yield $\beta=-3.0383\times10^{-4}$; the mismatch of $\sim2\%$ is
  likely due to the fact that $\alpha_1$ in Eq.~\eqref{eq_ev_pred} cannot be computed at the precise value of $c_0$ in the numerical setup. A similar agreement
  was also found in the vicinity of the other critical point at
$c_0 \approx 1.6709$.

As our final example, it is interesting to explore a case where
the relevant theory does not directly apply due to
limited regularity. As such an example, we consider an FPU model with the potential
of the form
\begin{equation}\label{eq:potentialat}
    V(u)=\begin{cases}
             \frac{u^2}{2}, & |u|\leq u_c \\
             \frac{\chi}{2} (|u|-u_c)^2 + u_c |u| - \frac{u_c^2}{2},  & |u|>u_c,
           \end{cases}
\end{equation}
which allows construction of explicit solitary waves \cite{at}, and $\Lambda(m)=0$; here $\chi>1$ and $u_c>0$. In this case the potential possesses only one continuous derivative, and hence the calculation of eigenvalues $\lambda$ and
FMs $\mu$ is less straightforward to justify, given the relevant
jump discontinuities. Nevertheless our detailed computations,
in line with the numerical results and stability conjecture in~\cite{at}, are in
a clear agreement with the criterion put forth analytically in this work. Namely,
$H'(c)>0$ in this case too corresponds to dynamical stability,
while $H'(c)<0$ leads to the manifestation of instability.

In order to qualitatively measure the instability, we have defined two diagnostic quantities. The first of them is the energy dispersion, given by
\[
\varepsilon(t)=1-\frac{\bar{H}(t)}{H},
\]
where $\bar{H}(t)$ is the energy at the nine central sites of the STW. In the case of a stable propagating wave, $\varepsilon(t)\sim10^{-4}$. The other quantity is the relative velocity change defined as
\[
\eta(t)=\frac{X(t)-X(0)}{tc}-1,
\] 
with $X(t)$ being the energy center of the STW.

\begin{figure}
\begin{center}
\begin{tabular}{cc}
\multicolumn{2}{c}{\includegraphics[width=7cm]{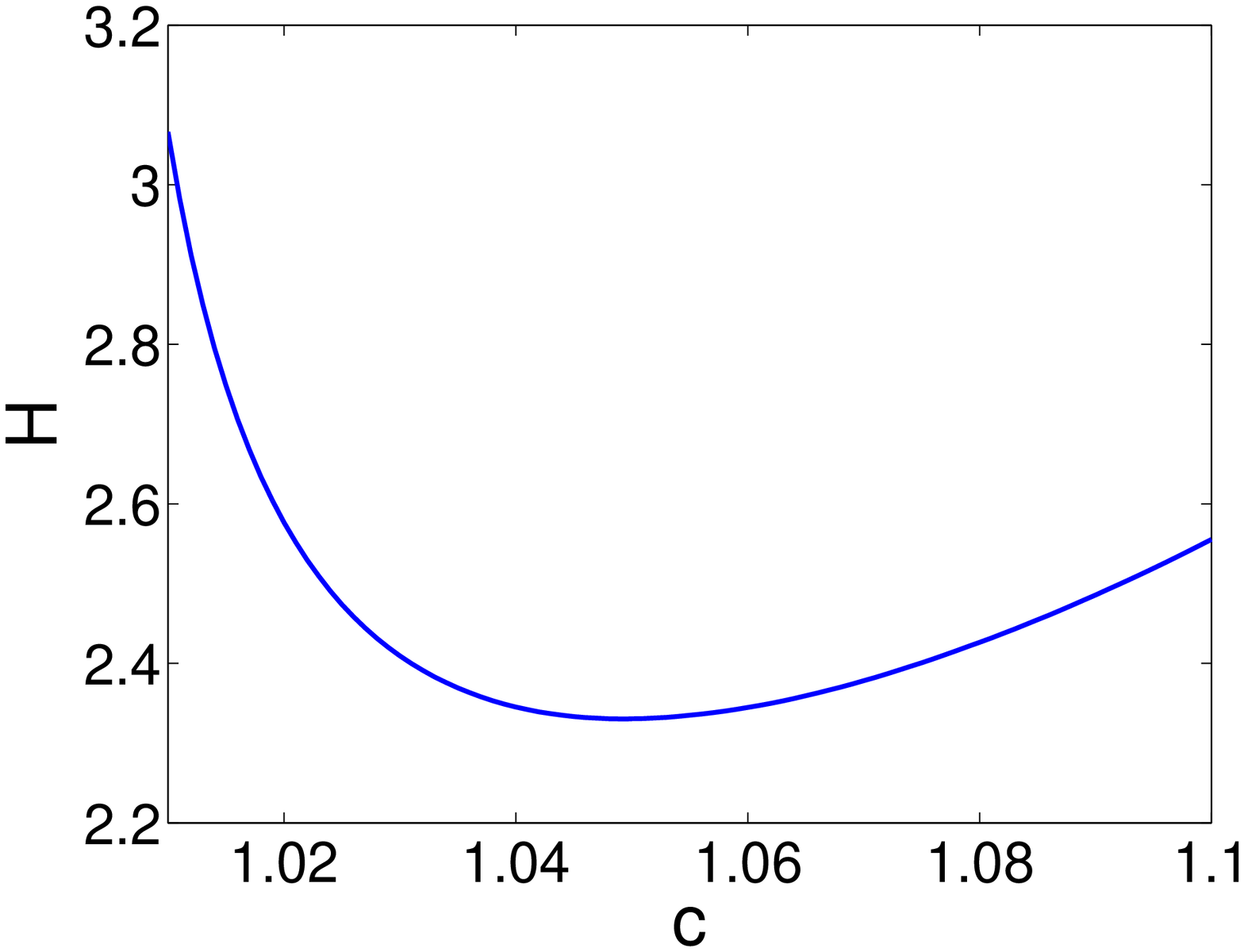}} \\
\includegraphics[width=7cm]{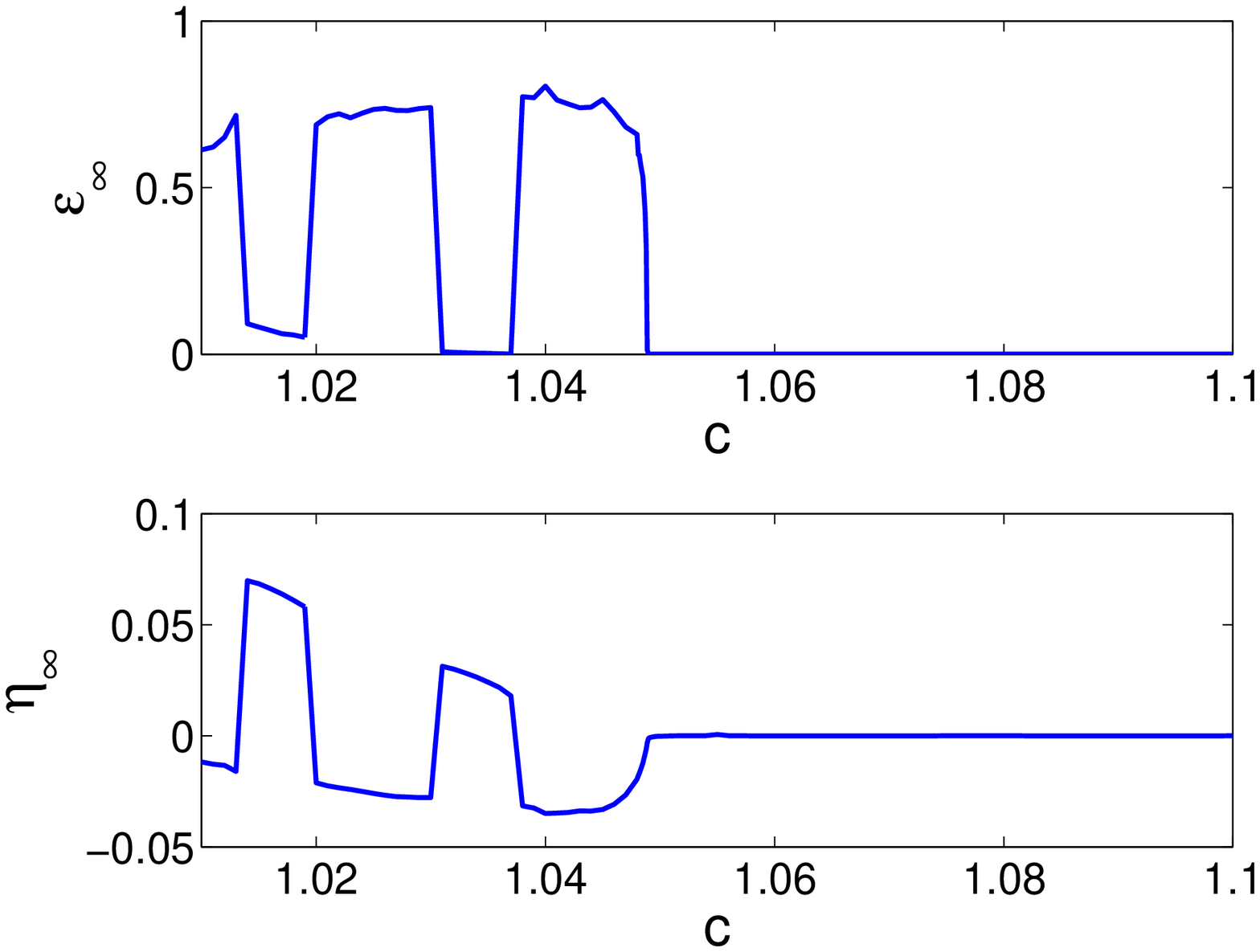} \\
\end{tabular}
\end{center}
\caption{Stability and instability of the lattice traveling waves of the model of~\cite{at} with the potential in Eq.~(\ref{eq:potentialat}). Here $\chi=4$ and $u_c=1$.
  The top panel displays the $H(c)$ dependence, which possesses a minimum at $c=c_0=1.0493$. The bottom panels show the dependence of the energy dispersion $\varepsilon_\infty$ and relative velocity change
  $\eta_\infty$ (see the text) with respect to the velocity $c$,
  which manifest the instability of solitary waves with $c<c_0$, where $c_0$ is such that
  $H'(c_0)=0$.}
\label{fig3}
\end{figure}

\begin{figure}
\begin{tabular}{c}
  \includegraphics[width=7cm]{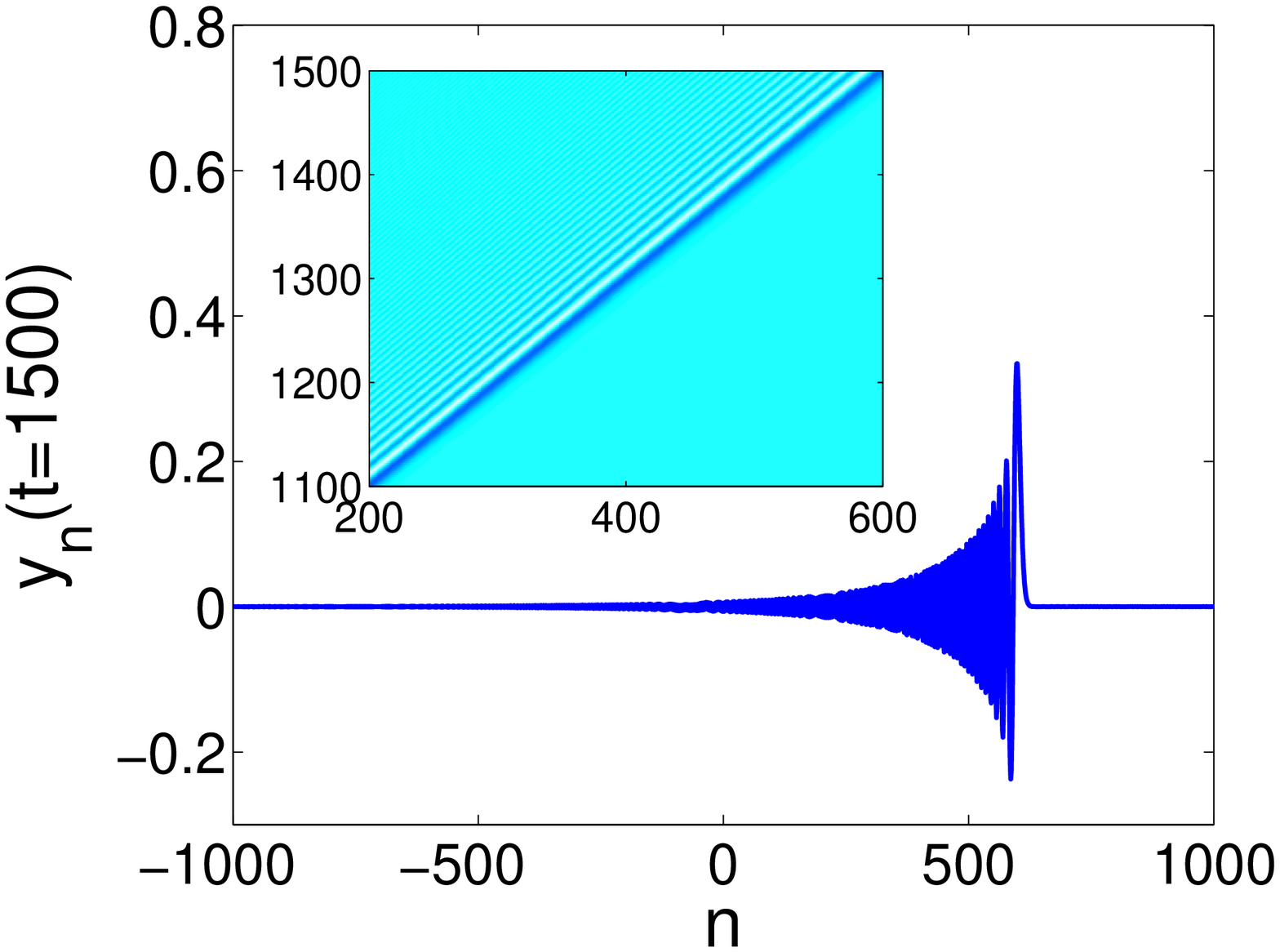} \\
  \includegraphics[width=7cm]{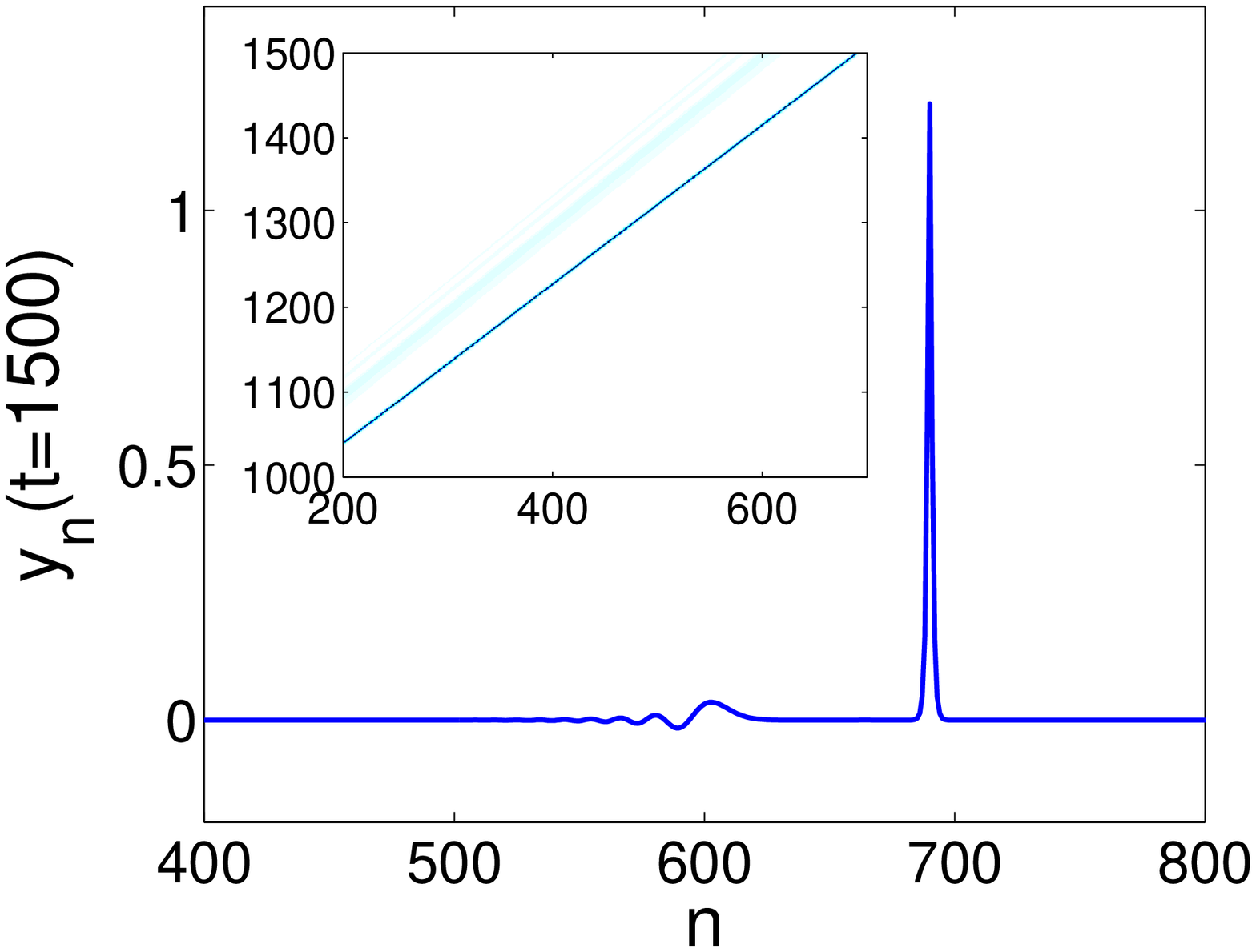} \\
\end{tabular}
\caption{Evolution of unstable travelling waves in the model of~\cite{at}
  with the potential in Eq.~(\ref{eq:potentialat}). Here $\chi=4$ and $u_c=1$.
  The panels show the profile of the strains $y_n(t)=u_{n+1}(t)-u_n(t)$ at $t=1500$ and zooms in
  the space-time evolution dynamics of the strains are represented in the insets. Top and bottom panels correspond to $c=1.025$ and $c=1.034$, respectively. In the the example shown in the bottom panel, the velocity eventually oscillates in time around an average value of 1.0626.}
\label{fig4}
\end{figure}

The top panel of Fig.~\ref{fig3} shows the curve $H(c)$ for the FPU model with
the potential of Eq.~(\ref{eq:potentialat}) and parameters $\chi=4$ and $u_c=1$; the bottom panels of this figure display the dependence of $\varepsilon_\infty\equiv\varepsilon(2000T)$ and $\eta_\infty\equiv\eta(2000T)$ with respect to $c$.
In accordance with our stability criterion, $\varepsilon_\infty\sim10^{-4}$ in the region for which $H'(c)>0$, confirming a stable propagation. In the region with $H'(c)<0$ there are three intervals of high dispersion, as measured
by corresponding values of $\varepsilon$,
and two intervals where the dispersion drops to low values.
The region of low dispersion corresponds to STWs whose velocity is higher
than the initial one (indeed, higher than the critical one and hence
reverting to the stable propagation regime). Fig. \ref{fig4} shows the evolution of unstable STWs in two cases, corresponding to high ($c=1.025$) and low ($c=1.034$) dispersion. In the former, linear waves are continuously being created and the STW degrades with time; in the latter, a linear wave is expelled from the STW, which transforms
into a wave with a different (now in the stable regime of $c>c_0$) velocity. Note that in addition to demonstrating instability of waves with $c<c_0$, these results suggest the potential bistability between dispersive waveforms and STWs with $c>c_0$.

\section{Conclusions and Future Challenges}
\label{sec:conclusions} 
In summary, in this
work we have presented a unified perspective connecting the stability
of lattice solitary traveling waves with that of discrete breathers of
an appropriate map involving running for the time associated with
moving by one lattice site and shifting back. We have also concisely
established a (sufficient but not necessary)
criterion for the change in spectral stability of the Hamiltonian
lattice STWs that seems to be in very good agreement
with numerical observations and to constitute a natural extension
of a criterion recently put forth for the spectral stability of
discrete breathers. The specific eigenvalue responsible
  for the instability was theoretically
  identified and favorably compared to detailed numerical computations.

Nevertheless, there are numerous problems that remain open for future
consideration.
One relevant issue  concerns the fact that the FM computation leads to as many multipliers
as lattice points, while the computation of eigenvalues for a STW involves a
partial differential equation (PDE). While the latter
  will capture the lattice instabilities, it may also
feature instabilities {\it absent} on the lattice, which
are a by-product of this PDE's ability to resolve scales
smaller than $h$. Hence, a more systematic connection between
the spectra of the two problems (and of the instabilities that
each may feature) is of paramount importance.
Observe also that while this work dealt with families of STWs parameterized by velocity, in some cases such entities
  occur for isolated velocity values~\cite{jcm,annav}, potentially
  being members of a wider family encompassing waveforms with non-vanishing
  tails. It would be
  interesting to explore whether our considerations can
be extended to such cases.
Another question is that of going to the continuum limit:
our proof did not directly use the underlying lattice nature of
the system (only its time reversal invariance). On the other hand,
in the continuum limit, symmetries (like Galilean or Lorentz
invariance) may arise. Future work will involve reconciling these two features in a
consistent continuum limit picture, as well as connecting our criterion with
well-established existing stability criteria, such as~\cite{vk,gss,bara}, in continuum systems.
Finally, analysis of the stability of lattice STWs in systems with limited regularity, such as our last example, also
merits future consideration.

\begin{acknowledgements}
J.C.-M. thanks financial support from MAT2016-79866-R project (AEI/FEDER, UE). A.V. acknowledges support by the U.S. National Science Foundation through the grant DMS-1506904. P.G.K.~gratefully acknowledges support
from the Alexander von Humboldt Foundation, the Greek Diaspora Fellowship
Program,
the US-NSF under grant PHY-1602994.
as well as the ERC under FP7, Marie
Curie Actions, People, International Research Staff Exchange
Scheme (IRSES-605096).
\end{acknowledgements}

\appendix

\section{Proof of the leading-order approximation of the near-zero eigenvalues}
\label{sec:details}
In this Appendix we prove Eq.~\eqref{eq_ev_pred} in Sec.~\ref{sec:criterion}, which provides the leading-order approximation of the eigenvalues splitting away from zero at velocities near the critical value $c_0$.

First, we observe that while we consider a lattice Hamiltonian system in the displacement form \eqref{eq0}, the problem can be alternatively formulated in terms of strain variables $y_n(t)=(S^{-1}-I)u_n(t)=u_{n+1}(t)-u_n(t)$, where we recall from Sec.~\ref{sec:floquet}
that $S$ denotes the shift operator such that $Su_n(t)=u_{n-1}(t)$. If the Hamiltonian energy density can be written as $\mathcal{H}(y,p,t)$, we have, for $Y=(y,p)^T$,
\begin{equation}
\label{eq1}
\dfrac{dY}{dt}= J_1 \nabla {\cal H}(Y), \quad J_1=\left(\begin{array}{cc}
0 & S^{-1}-I \\
I-S & 0
\end{array} \right).
\end{equation}
In what follows, we focus on the formulation (\ref{eq0}), but our arguments also work for Eq.~(\ref{eq1}).

Suppose Eq.~(\ref{eq0}) has a family of solitary traveling-wave solutions ${U}(t;c)$ parametrized by the velocity $c$ taking values in some continuous interval. Then
\[
U(t;c)=\hat{U}(\xi;c)=\left(\begin{array}{c}
  {\hat{u}(\xi)} \\
  {\hat{p}(\xi)}
  \end{array} \right), \quad \xi=n-ct,
\]
 where $\xi$ is the co-traveling frame variable and $\hat{p}(\xi;c)=-c\partial_{\xi}\hat{u}(\xi;c)$. Considering the ansatz
 \[
 U(\xi, t)=\left(\begin{array}{c}
  {\hat{u}(\xi)} \\
  {\hat{p}(\xi)}
\end{array} \right)+\epsilon e^{\lambda t}\left(\begin{array}{c}
  {W(\xi)} \\
  {P(\xi)}
\end{array} \right)=\hat{U}(\xi)+\epsilon e^{\lambda t}Z(\xi)
\]
with small $\epsilon$
and linearizing around the traveling wave $\hat{U}$, we obtain Eq.~\eqref{eq_ev},
where the operator $\mathcal{L}$ and its adjoint $\mathcal{L}^*$ are given by Eq.~\eqref{eq:L} and Eq.~\eqref{eq:Lstar}, respectively.

Suppose $\hat{U}\in H^1(\mathbb{R}^2)$, so that its partial derivatives in $\xi$ and $c$ are in $L^{2}(\mathbb{R}^2)$. While this assumption implies that the displacements are localized, for problems with kink-type traveling waves in terms of displacement that tend to nonzero constant limits at infinity, we can use the strain formulation \eqref{eq1}, in which case we assume that the traveling wave solution $\hat{Y}\in H^1(\mathbb{R}^2)$, i.e., the strains are localized. One can show that the operator $\mathcal{L}$ is densely defined on $L^{2}(\mathbb{R}^2)$.
By differentiating Eq.~(\ref{eq0}) in $\xi$ and $c$, respectively, we find that $\mathcal{L}e_0=0$ and $\mathcal{L}e_1=e_0$, where $e_0=-\partial_{\xi}\hat{U}$ and $e_1=\partial_c \hat{U}$ (or, more generally, $e_1=\partial_c \hat{U}+d_{10}e_0$, where $d_{10}$ is any constant), implying that the algebraic multiplicity of the eigenvalue $\lambda=0$ for $\mathcal{L}$ is at least two. Let $c_0$ denote the critical velocity such that $H'(c_0)=0$. Then $\langle e_1, J^{-1} e_0 \rangle=0$ at this critical value, and there exists $e_2$ such that $\mathcal{L}e_2=e_1$. Since
\[
\begin{split}
\langle e_2, J^{-1} e_0 \rangle&=\langle e_2, J^{-1}\mathcal{L}e_1 \rangle=\langle J^{-1}\mathcal{L}e_2, e_1 \rangle\\
&=\langle J^{-1}e_1, e_1 \rangle=0,
\end{split}
\]
we have $e_2 \in (\ker(\mathcal{L^*}))^{\perp}={\rm im}(\mathcal{L})$, so $e_2$ belongs to the range of $\mathcal{L}$, and hence there exists $e_3$ such that $\mathcal{L}e_3=e_2$. Assuming that the zero eigenvalue of $\mathcal{L}$ at $c_0$ is exactly quadruple, which is the generic case for traveling waves in Hamiltonian lattices due to symmetry, we have
\begin{equation}
\alpha_1=\langle e_0,J^{-1}e_3 \rangle=-\langle e_1,J^{-1}e_2 \rangle \neq 0.
\label{eq:alpha1}
\end{equation}

We now consider a neighborhood of the critical speed $c=c_0$ where the derivative $H'(c)$ changes
its sign. Assuming that $\hat{U}(\xi;c)$ is sufficiently smooth in $c$ near $c=c_0$, we have the expansion $\hat{U}(\xi;c_0+\epsilon)=U_0+\epsilon U_1+\epsilon^2 U_2+\dots$ for small enough $\epsilon$, where $U_0=\hat{U}(\xi;c_0)$, $U_1=(\partial_{c}\hat{U}(\xi;c))|_{c=c_0}$ and $U_2=\frac{1}{2}(\partial_{cc}\hat{U}(\xi;c))|_{c=c_0}$.
Accordingly, the operator $\mathcal{L}$ at $c=c_0+\epsilon$ can be written as $\mathcal{L}=\mathcal{L}_0+\epsilon\mathcal{L}_1+\epsilon^2\mathcal{L}_2+\dots$.
Let $\{e_0,e_1,e_2,e_3\}$ be the eigenfunction and generalized eigenfunctions of $\mathcal{L}_0$ for $\lambda=0$ such that
\[
\mathcal{L}_0^3e_3=\mathcal{L}_0^2e_2=\mathcal{L}_0 e_1=e_0=-\partial_{\xi}\hat{U}(\xi;c_0).
\]
We then define the following constants:
\begin{equation}
K_{jk}=\langle J^{-1}e_j, \mathcal{L}_1 e_k \rangle, \quad L_{jk}=\langle J^{-1}e_j, \mathcal{L}_2 e_k \rangle.
\label{eq:KL}
\end{equation}

\begin{remark}
If the generalized kernel of $\mathcal{L}_0$ is exactly four-dimensional, then only the cases $\lambda\sim \epsilon^{1/2}$ and $\lambda\sim \epsilon$ are possible.
\end{remark}
\noindent Indeed, this follows from the fact that two of the four eigenvalues of $\mathcal{L}$ are always zero. It suffices to calculate the leading-order terms of the eigenvalues for the perturbed operator $\mathcal{L}$ at $c=c_0+\epsilon$. By restricting the operator in the invariant subspace $G_4=\text{cl}({\rm span}\{e_0,e_1,e_2,e_3\})$, the question reduces to the perturbation of the matrix
\[
A=\left(\begin{array}{cccc}
  0 & 0 & 0  & 0\\
  1 & 0 & 0 & 0\\
  0 & 1 & 0 & 0 \\
  0 & 0 & 1 & 0
\end{array} \right)
\]
with two constraints that hold for any $c$,
\begin{equation}
\mathcal{L}(\partial_{\xi} \hat{U}(\xi;c))=0
\label{eq:constraint1}
\end{equation}
and
\begin{equation}
\mathcal{L}(\partial_{c} \hat{U}(\xi;c))=-(\partial_{\xi} \hat{U}(\xi;c)).
\label{eq:constraint2}
\end{equation}
 Note that the characteristic polynomial of the unperturbed matrix $A$ is $\lambda^4=0$. For the matrix $A$ with $O(\epsilon)$ perturbation, the characteristic polynomial is $\lambda^4+a_3 \lambda^3+a_2 \lambda^2+a_1\lambda +a_0=0$ where the coefficients $a_j$ are at most $O(\epsilon)$. Moreover, due to two existing constraints in Eq.~\eqref{eq:constraint1} and Eq.~\eqref{eq:constraint2}, two of the eigenvalues are always zero, so we have $\lambda^2(\lambda^2+a_3 \lambda+a_2)=0$. Thus, either $\lambda\sim \epsilon^{1/2}$ (if $a_2\neq 0$) or $\lambda\sim\epsilon$ (if $a_2=0$).

Here we focus on the case $\lambda\sim \epsilon^{1/2}$ and show below that it requires $H''(c_0)\neq 0$. Since Eq.~\eqref{eq:constraint1} holds for any $c$, direct calculation shows that
\begin{eqnarray}
\label{eqn_U_expand1_1}
0&=&\mathcal{L}_0 (\partial_{\xi} U_0),\\
\label{eqn_U_expand1_2}
0&=&\mathcal{L}_0 (\partial_{\xi} U_1)+\mathcal{L}_1 (\partial_{\xi} U_0),\\
\label{eqn_U_expand1_3}
0&=&\mathcal{L}_0 (\partial_{\xi} U_2)+\mathcal{L}_1 (\partial_{\xi} U_1)+\mathcal{L}_2 (\partial_{\xi} U_0).
\end{eqnarray}
Moreover, utilizing the fact that \eqref{eq:constraint2} is true for any $c$, one can expand both sides in $\epsilon$ and obtain
\begin{eqnarray}
\label{eqn_U_expand2_1}
\mathcal{L}_0 U_1&=&-\partial_{\xi}U_0, \\
\label{eqn_U_expand2_2}
2\mathcal{L}_0 U_2 + \mathcal{L}_1 U_1 &=& -\partial_{\xi}U_1.
\end{eqnarray}
Since $H'(c_0)=0$, we can write $H(c)=H(c_0)+\frac{\epsilon^2}{2}H''(c_0)+o(\epsilon^2)$, where
\begin{equation}
\begin{split}
&H''(c_0)=\nabla H \cdot 2U_2 + U_1 \cdot  \nabla^2 H U_1\\
 &= \int [c_0 J^{-1}(e_0) \cdot 2 U_2
  + U_1\cdot J^{-1} (\mathcal{L}_0- c_0\partial_{\xi}) U_1 ] d\xi\\
&= \int [c_0 J^{-1} \mathcal{L}_0 U_1 \cdot 2 U_2
  + U_1\cdot (J^{-1}\mathcal{L}_0 U_1- c_0 J^{-1}\partial_{\xi} U_1) ] d\xi\\
&=  \int U_1 \cdot (J^{-1}e_0) d\xi+ c_0\int U_1 \cdot J^{-1}(\mathcal{L}_0 2 U_2 -\partial_{\xi} U_1) d\xi\\
&= c_0\int U_1 \cdot J^{-1}(-2\partial_{\xi}U_1 -\mathcal{L}_1 U_1) d\xi\\
&=  -c_0\int \mathcal{L}_0 e_2 \cdot J^{-1}2\partial_{\xi}U_1 d\xi- c_0 \int e_1 \cdot J^{-1}\mathcal{L}_1 e_1 d\xi\\
&=  -c_0\int  e_2 \cdot (-1)J^{-1}\mathcal{L}_0 2\partial_{\xi}U_1 d\xi - c_0 \int e_1 \cdot J^{-1}\mathcal{L}_1 e_1 d\xi\\
&=  c_0\int  e_2 \cdot J^{-1}\mathcal{L}_1 2 e_0 d\xi - c_0 \int e_1 \cdot J^{-1}\mathcal{L}_1 e_1 d\xi\\
&=  -c_0(2K_{20}-K_{11}).
\end{split}
\end{equation}
Assuming $\lambda=\epsilon^{1/2}\lambda_1+\epsilon \lambda_2+\epsilon^{3/2}\lambda_3+\dots$ and $Z=Z_0+\epsilon^{1/2} Z_1+\epsilon Z_2+\epsilon^{3/2} Z_3+\dots$ and substituting these into Eq.~(\ref{eq_ev}), we obtain
\beq
\label{eqn_lambda_expand0}
0=\mathcal{L}_0 Z_0,
\eeq
\beq
\label{eqn_lambda_expand1}
\lambda_1 Z_0=\mathcal{L}_0 Z_1,
\eeq
\beq
\label{eqn_lambda_expand2}
\lambda_1 Z_1+\lambda_2 Z_0=\mathcal{L}_0 Z_2+\mathcal{L}_1 Z_0,
\eeq
\beq
\label{eqn_lambda_expand3}
\lambda_1 Z_2+\lambda_2 Z_1+\lambda_3 Z_0=\mathcal{L}_0 Z_3+\mathcal{L}_1 Z_1,
\eeq
\beq
\label{eqn_lambda_expand4}
\begin{split}
\lambda_1 Z_3+\lambda_2 Z_2+\lambda_3 Z_1+\lambda_4 Z_0&=\mathcal{L}_0 Z_4+\mathcal{L}_1 Z_2
\\&+\mathcal{L}_2 Z_0.
\end{split}
\eeq
From Eq.~(\ref{eqn_lambda_expand0}), we find that $Z_0=-\partial_{\xi}\hat{U}(\xi;c_0)=e_0$. Then Eq.~(\ref{eqn_lambda_expand1}) suggests that $Z_1=\lambda_1 e_1 + d_{10} e_0$, where $d_{10}$ is a constant. Note that $Z_2$ and $Z_3$ can be written as
\[
Z_2=\sum_{j=0}^3 (d_{2j}e_j) + Z_2^{\perp}, \quad Z_3=\sum_{j=0}^3 (d_{3j}e_j) + Z_3^{\perp},
\]
where $Z_2^{\perp}$ and $Z_3^{\perp}$ are in $G_4^{\perp}$, and $d_{2j}$, $d_{3j}$, $j=0,\dots,3$ are constants. Projecting Eq.~(\ref{eqn_lambda_expand2}) onto $J^{-1} e_0$ yields
$\lambda_1 \langle J^{-1} e_0, Z_1 \rangle +\lambda_2 \langle J^{-1} e_0, Z_0 \rangle = K_{00} $. The left hand side is zero since $H'(c_0)=0$, and one can show that the right hand side vanishes ($K_{00}=0$) upon considering Eq.~(\ref{eqn_U_expand1_2}).
Projecting Eq.~(\ref{eqn_lambda_expand2}) onto $J^{-1} e_1$ and recalling Eq.~\eqref{eq:alpha1}, we obtain $d_{23}\langle J^{-1}e_1, e_2 \rangle+K_{10}=d_{23}\alpha_1+K_{10}=0$, so
\beq
d_{23}=-\frac{K_{10}}{\alpha_1}.
\label{eq:d23}
\eeq
Projecting Eq.~(\ref{eqn_lambda_expand2}) onto $J^{-1} e_2$ and using \eqref{eq:alpha1}, we find that $d_{22}\langle J^{-1}e_2, e_1 \rangle+K_{20}=-d_{22}\alpha_1+K_{20}=-\lambda_1^2 \alpha_1$,
and thus
\beq
d_{22}=\lambda_1^2+\frac{K_{20}}{\alpha_1}.
\label{eq:d22}
\eeq
Projecting Eq.~(\ref{eqn_lambda_expand2}) onto $J^{-1} e_3$, we have
\beq
d_{21}\alpha_1+d_{23}\alpha_2+K_{30}= (d_{10}\lambda_1+\lambda_2)\alpha_1,
\label{eq:d10}
\eeq
where we used Eq.~\eqref{eq:alpha1} and set $\alpha_2=\langle J^{-1}e_3,e_2\rangle$.
Projecting Eq.~(\ref{eqn_lambda_expand3}) onto $J^{-1} e_0$ yields
$\lambda_1 d_{23} \langle J^{-1} e_0, e_3 \rangle=-\lambda_1 d_{23} \alpha_1 = \lambda_1 K_{01}$, which again yields Eq.~\eqref{eq:d23} since $K_{01}=K_{10}$.
Projecting Eq.~(\ref{eqn_lambda_expand3}) onto $J^{-1} e_1$, we obtain
\beq
d_{33}\alpha_1+\lambda_1 K_{11}+d_{10}K_{10}=\lambda_1 d_{22} \alpha_1.
\label{eq:d33}
\eeq
Finally, projection of Eq.~(\ref{eqn_lambda_expand4}) onto $J^{-1} e_0$ yields
\beq
\begin{split}
d_{21}K_{01}&+d_{22}K_{02}+d_{23}K_{03}+L_{00}\\
&=-\alpha_1 (\lambda_1 d_{33}+\lambda_2 d_{23}).
\end{split}
\label{eq:d21}
\eeq

Using the equations \eqref{eq:d23}, \eqref{eq:d22}, \eqref{eq:d10}, \eqref{eq:d33}, \eqref{eq:d21} along with the fact that
$K$ is symmetric, we obtain
\[
\begin{split}
\alpha_1 \lambda_1^4 &+ (2 K_{20}-K_{11})\lambda_1^2\\
&+(L_{00}+\frac{K_{20}^2}{\alpha_1}+\frac{\alpha_2 K_{10}^2}{\alpha_1^2}-\frac{2 K_{10}K_{30}}{\alpha_1})=0.
\end{split}
\]
Since two eigenvalues are always zero, this equation should have two zero roots. This implies
\[
L_{00}+\frac{K_{20}^2}{\alpha_1}+\frac{\alpha_2 K_{10}^2}{\alpha_1^2}-\frac{2 K_{10}K_{30}}{\alpha_1}=0,
\]
which can also be shown directly using projections of Eq.~\eqref{eqn_U_expand1_2} onto $J^{-1}e_j$, $j=0,\dots,3$, and projection of Eq.~\eqref{eqn_U_expand1_3} onto $J^{-1}e_0$. We then obtain
\[
\lambda_1^2=-\frac{2K_{20}-K_{11}}{\alpha_1}=\frac{H''(c_0)}{\alpha_1 c_0}.
\]
Thus, for $\lambda \sim \epsilon^{1/2}$ it is necessary to have $H''(c_0) \neq 0$, and the behavior of the two eigenvalues splitting away from zero at $c \neq c_0$ is described by Eq.~\eqref{eq_ev_pred} in Sec.~\ref{sec:criterion}.

\section{Numerical methods for computing solitary traveling waves}
\label{sec:numer}
In this Appendix, we describe the numerical procedures we used to compute solitary waves in a lattice with the Hamiltonian in Eq.~\eqref{eq:Ham} and analyze their stability. The governing equations corresponding to Eq.~\eqref{eq:Ham} are
\begin{equation}\label{eq:dyn1}
\begin{split}
    \ddot u_n&-V'(u_{n+1}-u_n)+V'(u_n-u_{n-1})\\
    &+\sum_{m=1}^\infty\Lambda(m)(2u_n-u_{n+m}-u_{n-m})=0,
    \end{split}
\end{equation}
where the overdots here and in what follows denote the
time derivatives.
 Since the solitary solutions we consider are kink-like in terms of displacement, it is more convenient to rewrite Eq.~(\ref{eq:dyn1}) in terms of the strain variables $y_n=u_{n+1}-u_{n}$, obtaining
\begin{equation}\label{eq:dyn}
\begin{split}
   \ddot y_n&+2V'(y_n)-V'(y_{n+1})-V'(y_{n-1})\\
   &+\sum_{m=1}^\infty\Lambda(m)(2y_n-y_{n+m}-y_{n-m})=0.
   \end{split}
\end{equation}

To find solitary traveling wave solutions, we use the procedure followed in \cite{Yasuda}. To this end, we seek solutions of Eq.~(\ref{eq:dyn}) in the co-traveling frame corresponding to velocity $c$:
\[
y_n(t)=\Phi(\xi,t), \quad \xi=n-ct,
\]
obtaining the advance-delay partial differential equation
\begin{equation}\label{eq:dynade}
\begin{split}
&\Phi_{tt}+c^2\Phi_{\xi\xi}-2c\Phi_{\xi t}\\
&=V'(\Phi(\xi+1,t))+V'(\Phi(\xi-1,t))-2V'(\Phi(\xi,t))\\
&-\sum_{m=1}^\infty\Lambda(m)(2\Phi(\xi,t)-\Phi(\xi+m,t)-\Phi(\xi-m,t)).
\end{split}
\end{equation}
Traveling waves $\phi(\xi)$ are stationary solutions of Eq.~(\ref{eq:dynade}). They satisfy the advance-delay differential equation
\begin{equation}\label{eq:ade}
\begin{split}
    &c^2\phi''(\xi)+2V'(\phi(\xi))-V'(\phi(\xi+1))-V'(\phi(\xi-1))\\
    &+\sum_{m=1}^\infty\Lambda(m)(2\phi(\xi)-\phi(\xi+m)-\phi(\xi-m))=0.
    \end{split}
\end{equation}
Solitary traveling waves are solutions that in addition satisfy
\beq
\lim_{\xi\rightarrow \pm \infty}\phi(\xi)=0.
\label{eq:BCs}
\eeq
Following the approach in \cite{eilbeck}, we assume that $\phi(\xi)=o(1/\xi)$ and $\phi'(\xi)=o(1/\xi^2)$ as $|\xi| \rightarrow \infty$, multiply Eq.~(\ref{eq:ade}) by $\xi^2$ and integrate by parts to derive the identity
\begin{equation}\label{eq:ade1}
    \left[c^2-\sum_{m=1}^{\infty}m^2\Lambda(m)\right]\int_{-\infty}^{\infty} \phi(\xi)\mathrm{d}\xi-\int_{-\infty}^{\infty} V'(\phi(\xi))\mathrm{d}\xi=0,
\end{equation}
which imposes the constraint \eqref{eq:BCs} on the traveling wave solutions. Here we assume that $\Lambda(m)$ decays faster than $1/m^3$ at infinity, so that the series on the left hand side converges.

To solve Eq.~(\ref{eq:ade}) numerically, we introduce a discrete mesh with step $\Delta\xi$, where $1/\Delta\xi$ is an integer, so that the advance and delay terms $\phi(\xi \pm m)$ are well defined on the mesh. We then use a {Fourier} spectral collocation method for the resulting system with periodic boundary conditions \cite{Trefethen} with large period $L$. Implementation of this method requires an even number $\mathcal{N}$ of collocation points $\xi_j\equiv j\Delta\xi$, with $j=-\mathcal{N}/2+1,\ldots,\mathcal{N}/2$, yielding a system for $\xi$ in the domain $(L/2,L/2]$, with $L=\mathcal{N}\Delta\xi$ being an even number, and the long-range interactions are appropriately truncated.
  To ensure that the solutions satisfy Eq.~\eqref{eq:BCs}, we additionally impose
  a trapezoidal approximation of Eq.~\eqref{eq:ade1} on the truncated interval at the collocation points. This procedure is independent of the potential and the interaction range. However, the choices of $\Delta\xi$ and $L$ depend on the nature of the problem. In the particular cases considered in the paper, we used  $\Delta\xi=0.1$, $L=800$ for the $\alpha$-FPU lattice with nearest-neighbor potential in Eq.~\eqref{eq:FPU} and Kac-Baker long-range interactions with coefficients in Eq.~\eqref{eq:KB} and $\Delta\xi=0.025$, $L=200$ for the FPU lattice with piecewise quadratic short-range interaction potential in Eq.~\eqref{eq:potentialat} and no long-range interactions.

To investigate spectral stability of an obtained traveling wave $\phi(\xi)$, we substitute
\[
\Phi(\xi,t)=\phi(\xi)+\epsilon a(\xi)\exp(\lambda t),
\]
into Eq.~\eqref{eq:dynade} and consider $O(\epsilon)$ terms resulting from this perturbation. This yields the following quadratic eigenvalue problem:
\begin{equation}\label{eq:perturbade}
\begin{split}
\lambda^2a(\xi)&=-c^2a''(\xi)+2\lambda c a'(\xi)
    -2V''(\phi(\xi))a(\xi)\\
    &+V''(\phi(\xi+1))a(\xi+1)+V''(\phi(\xi-1))a(\xi-1)\\
    &-\sum_{m=1}^\infty\Lambda(m)(2a(\xi)-a(\xi+m)-a(\xi-m)).
\end{split}
\end{equation}
By defining $b(\xi)=\lambda a(\xi)$, we transform this equation into the regular eigenvalue problem
\begin{equation}
\lambda\left(
\begin{array}{c}
  a(\xi) \\
  b(\xi)
\end{array}
\right)
=
\mathcal{M}
\left(
\begin{array}{c}
  a(\xi) \\
  b(\xi)
\end{array}
\right)
\end{equation}
for the corresponding linear advance-delay differential operator $\mathcal{M}$. Note that this problem is equivalent to the eigenvalue problem (\ref{eq_ev}) via the transformation $(a(\xi), b(\xi))=(W(\xi), P(\xi)+c W'(\xi))$. Spectral stability can be determined by analyzing the spectrum of the operator $\mathcal{M}$ after discretizing the eigenvalue problem the same way as the nonlinear Eq.~\eqref{eq:ade} and again using periodic boundary conditions. A solution is stable when the spectrum contains no real eigenvalues.

An alternative method for determining the stability of the traveling waves is to use Floquet analysis. To this end, we cast traveling waves $\phi(\xi)$ as fixed points of the map
\begin{equation}\label{eq:map}
\left[\begin{array}{c}
  \{y_{n+1}(T)\} \\ \{\dot y_{n+1}(T)\} \\  \end{array}\right]
  \rightarrow
  \left[\begin{array}{c}
  \{y_{n}(0)\} \\ \{\dot y_{n}(0)\} \\  \end{array}\right],
\end{equation}
which is periodic modulo shift by one lattice point, with period $T=1/c$. Indeed, one easily checks that $\hat{y}_n(t)=\phi(n-ct)=\phi(n-t/T)$ satisfies $\hat{y}_{n+1}(T)=\hat{y}_n(0)=\phi(n)$ and $\dot{\hat y}_{n+1}(T)=\dot{\hat y}_{n}(0)=-c\phi'(n)$.  To apply the Floquet analysis, we trace time evolution of a small perturbation $\epsilon w_n(t)$ of the periodic-modulo-shift (traveling wave) solution $\hat{r}_n(t)$. This perturbation is introduced in Eq.~(\ref{eq:dyn}) via $y_n(t)=\hat{y}_n(t)+\epsilon w_n(t)$. The resulting $O(\epsilon)$ equation reads
\begin{equation}\label{eq:perturb}
\begin{split}
    &\ddot w_n+2V''(\hat{y}_n)w_n-V''(\hat{y}_{n+1})w_{n+1}-V''(\hat{y}_{n-1})w_{n-1}\\
    &+\sum_{m=1}^\infty\Lambda(m)(2w_n-w_{n+m}-w_{n-m})=0.
    \end{split}
\end{equation}
Then, in the framework of Floquet analysis, the stability properties of periodic orbits are resolved by diagonalizing the monodromy matrix $\mathcal{F}$ (representation of the Floquet operator for finite systems), which is defined as:
\begin{equation}\label{eq:Floquet}
\left[\begin{array}{c}
  \{w_{n+1}(T)\} \\ \{\dot w_{n+1}(T)\} \\  \end{array}\right]
  =\mathcal{F}
  \left[\begin{array}{c}
  \{w_{n}(0)\} \\ \{\dot w_{n}(0)\} \\  \end{array}\right] .
\end{equation}
For the symplectic Hamiltonian systems considered in this work, the linear stability of the solutions requires that the monodromy eigenvalues $\mu$ (also called Floquet multipliers) lie on the unit circle. The Floquet multipliers can thus
be written as $\mu=\exp(i\theta)$, with Floquet exponent $\theta$.

Note that the two procedures for analyzing spectral stability described above require the potential $V(u)$ to be twice differentiable, as in the case of the $\alpha$-FPU problem considered in Sec.~\ref{sec:examples}. Due to the absence of such regularity in the case of the piecewise quadratic potential in Eq.~\eqref{eq:potentialat}, the examination
  of stability was performed solely on the basis
  of direct numerical simulations. Specifically, it was analyzed
  by means of tracking the dynamics of a slightly perturbed solution $[\{\hat{y}_{n}(0)\},\{\dot{\hat y}_{n}(0)\}]$. To this aim, the fourth order explicit and symplectic Runge-Kutta-Nystr\"om method developed in \cite{Calvo}, with time step equal to $10^{-3}$, was used.


\begin{thebibliography}{99}

\bibitem{FPU}
E. Fermi, J. Pasta, and S. Ulam, Tech. Rep. Los Alamos Nat.
Lab. LA1940 (1955);
  D.~K. Campbell, P. Rosenau, and G.~M.
Zaslavsky, Chaos {\bf 15}, 015101 (2005);
G. Galavotti (Ed.) {\it The
Fermi-Pasta-Ulam Problem: A Status Report} (Springer-Verlag,
New York, 2008).


\bibitem{toda} M. Toda, {\it Theory of nonlinear lattices}, Springer-Verlag
(Berlin, 1989).

\bibitem{nesterenko}  V.~F. Nesterenko, {\it Dynamics of Heterogeneous Materials}, Chapter 1, Springer-Verlag (New York, 2001).


\bibitem{sen} S. Sen, J. Hong, J. Bang,  E. Avalos, R. Doney,
Phys. Rep. {\bf 462}, 21-66 (2008).


\bibitem{review} C. Chong, M.~A. Porter, P.~G. Kevrekidis, C. Daraio,
  arXiv:1612.03977.


\bibitem{mertens} D. Hochstrasser, F.~G. Mertens, and H. B{\"u}ttner, Physica D
  {\bf 35}, 259 (1989).

\bibitem{eilbeck} J.~C. Eilbeck, R. Flesch,
  Phys. Lett. A{\bf 149}, 200 (1990).

\bibitem{english} J.~M. English, and R.~L. Pego,
  Proc. Am. Math. Soc. {\bf 133}, 1763 (2005).


\bibitem{wayne} G.~N. Benes, A. Hoffman, and C.~E. Wayne.
  J.
Math. Anal. Appl. {\bf 386}, 445 (2012).

\bibitem{pegof3} G. Friesecke, R.~L. Pego,
  Nonlinearity {\bf 17}, 207 (2004).

\bibitem{pegof4} G. Friesecke, R.~L. Pego,
  Nonlinearity {\bf 15}, 1343 (2002).

\bibitem{mertens1} S.~F. Mingaleev, Y.~B. Gaididei,
  F.~G. Mertens, Phys. Rev. E {\bf 58}, 3833 (1998).

\bibitem{mertens2} S.~F. Mingaleev, Y.~B. Gaididei,
  F.~G. Mertens, Phys. Rev. E {\bf 61}, R1044 (2000).

\bibitem{at} L. Truskinovsky, A. Vainchtein,
  Phys. Rev. E {\bf 90}, 042903 (2014).

\bibitem{floria}  J. G{\'o}mez-Garde\~{n}es, F. Falo, and L. M. Floria, Phys. Lett. A {\bf 332}, 213-219 (2004).

\bibitem{dmp} P.~G. Kevrekidis, J. Cuevas-Maraver,
  D.E. Pelinovsky, Phys. Rev. Lett. {\bf 117}, 094101 (2016).

\bibitem{aubry} S. Aubry, Physica D {\bf 103}, 201 (1997).

\bibitem{FlachPR2008} S.~Flach and A.~V.~Gorbach, Phys. Rep. {\bf 467}, 1 (2008).


\bibitem{arnold} V.~I. Arnold,
  {\it Mathematical Methods of Classical Mechanics},
  Springer-Verlag (New York, 1989).

\bibitem{ourjesus} J. Cuevas, V. Koukouloyannis, P.~G. Kevrekidis,
  J.F.R. Archilla,
Int. J. Bif. Chaos {\bf 21}, 2161 (2011).


\bibitem{mizu} T. Mizumachi, R.~L. Pego,
Nonlinearity {\bf 21}, 2099 (2008).

\bibitem{jphysa} H. Xu, P.~G. Kevrekidis, and A. Stefanov, J. Phys. A  {\bf 48}, 195204 (2015).


\bibitem{jcm} T.~R.~O. Melvin, A.~R. Champneys, P.~G. Kevrekidis, and J. Cuevas,
Phys. Rev. Lett. {\bf 97}, 124101 (2006).

\bibitem{annav} A. Vainchtein, Y. Starosvetsky,
  J.~D. Wright, and R. Perline,
Phys. Rev. E {\bf 93}, 042210 (2016).

\bibitem{gss} M. Grillakis, J. Shatah, W. Strauss,
J. Funct. Anal. {\bf 74} 1, 160 (1987).

\bibitem{vk} N.~G. Vakhitov, A.~A. Kolokolov,
Radiophys. Quantum Electron. {\bf 16}, 783 (1973).

\bibitem{bara} I.~V. Barashenkov,
Phys. Rev. Lett. {\bf 77}, 1193 (1996)

\bibitem{Yasuda} H. Yasuda, C. Chong, E.~G. Charalampidis, P.~G. Kevrekidis and J. Yang. Phys. Rev. E {\bf 90}, 043004 (2016).

\bibitem{Trefethen} L.~N. Trefethen, {\em Spectral methods in MATLAB}. SIAM, Philadelphia (2000).

\bibitem{Calvo} M.~P. Calvo and J.~M. Sanz Serna, SIAM J. Sci. Comput. {\bf} 14, 936 (1993).

\end{thebibliography}
\end{document}